\pdfoutput=1 
\documentclass{JINST}
\usepackage{float}
\usepackage[skip=0pt]{caption}
\usepackage[font=small]{subcaption}
\usepackage{lineno}
\usepackage{upgreek}
\newcommand{\pb}[1]{\parbox{0.42\linewidth}{#1} \vspace{0pt}}
\title{Diamond pixel detector for beam profile monitoring in COMET experiment at
J-PARC}

\author{M. \v{C}erv$^a$,
P. Sarin$^b$\thanks{Corresponding author.}, 
H. Pernegger$^a$,
P. Vageeswaran$^b$~and
E. Griesmayer$^c$\\
\llap{$^a$}CERN PH,
Geneva, Swizerland\\
\llap{$^b$}Indian Institute of Technology,
Bombay, India\\
\llap{$^c$}Vienna University of Technology,
Vienna, Austria\\
E-mail: \email{pradeepsarin@iitb.ac.in}}

\abstract{
We present the design and initial prototype results of a pixellized proton 
beam profile monitor for the COMET experiment at \mbox{J-PARC}. 
The goal of COMET is to look for charged lepton flavor 
violation by direct $\mu$ to e conversion at a sensitivity of $10^{-19}$. 
An 8 GeV proton beam pulsed at 100 ns with $10^{10}$ protons/s will 
be used to create muons through pion production and decay. 
In the final experiment, the proton flux will be raised to $10^{14}$ 
protons/sec to increase the sensitivity. 
These requirements of harsh radiation tolerance and fast readout make 
diamond a good choice for constructing a beam profile monitor in COMET.
We present first results of the characterization of single crystal diamond (scCVD) 
sourced from a new company, \textsc{2a systems} Singapore. 
Our measurements indicate excellent charge collection efficiency
 and high carrier mobility down to cryogenic temperatures.
}


\begin{document}

\section{Motivation and test setup}\label{sec:setup}

Diamond has emerged as an attractive alternative to
silicon for detection of ionizing radiation. Properties that make 
diamond a superior sensor material are summarized in Table \ref{tab:c-prop}.
Diamond detector structures require no dopants 
and can be operated at higher fields for faster charge collection.
By far the most attractive property of diamond is its high radiation tolerance.
\begin{table}[htbp]
\caption{Characteristics of diamond compared with silicon}
\label{tab:c-prop}
\centering
\begin{tabular}{|llll|}
\hline
Property(units)&Diamond&Silicon&Benefit\\
\hline
\hline
Bandgap $(eV)$&$5.5$&$1.12$&Low thermal carrier density\\
\hline
Breakdown field $(V/cm)$&$10^7$&$10^6$&High field operation\\
\hline
Mean energy to create e-h pair $(eV)$&$13$&$3.6$ & \\
\hline
Dielectric constant & 5.7 & 11.9 & Small detector capacitance, less noise \\
\hline
Displacement energy (eV/atom) & 43 & 13-20 & High radiation tolerance \\
\hline 
\end{tabular}
\end{table}

This paper presents first results of the study of single crystal diamond
properties from a new supplier: \textsc{2a systems}, Singapore.
The sample received from \textsc{2a systems} (\textsc{2a}) is 
460~$\upmu m$ thick with a cross-section area of $5.3mm\times 5.3mm$. 
For comparison, we have used a second sample from \textsc{element6} (\textsc{e6})
of comparable area
and thickness 538$\upmu m$. \textsc{2a} is metallized with 
50 nm Cr $+$ 200 nm Au ohmic contacts on both sides. 
DC tests on \textsc{2a} indicate a capacitance of 2.2~pF at 
the nominal operating voltage of $1V/\upmu m$.

We performed two types of tests to determine the signal characteristics in
diamond:
\begin{enumerate}
\item We measure the current induced as a function of time by a stopping 
$\alpha$ particle with the transient current 
technique (TCT)\cite{tctref}. This is done as a function of 
temperature down to $4~K$. 
\item Charge collection efficiency of signal deposited by minimum ionizing
$\beta$ particles is measured as a function of the bias voltage. 
\end{enumerate}

\begin{figure}[!htbp]
  \centering
  \begin{subfigure}[t]{0.5\textwidth}
  \includegraphics[width=0.9\textwidth]{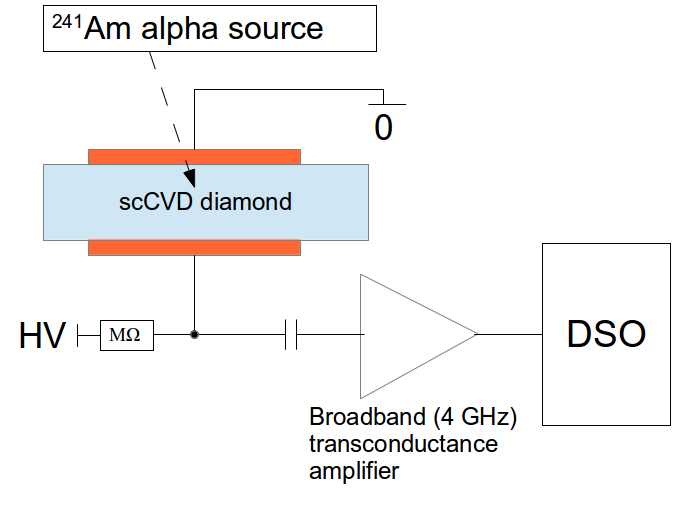}
  \caption{Setup for TCT measurements. DSO triggers on the signal}
  \label{fig:tct}
  \end{subfigure}~
  \begin{subfigure}[t]{0.5\textwidth}
  \includegraphics[width=0.9\textwidth]{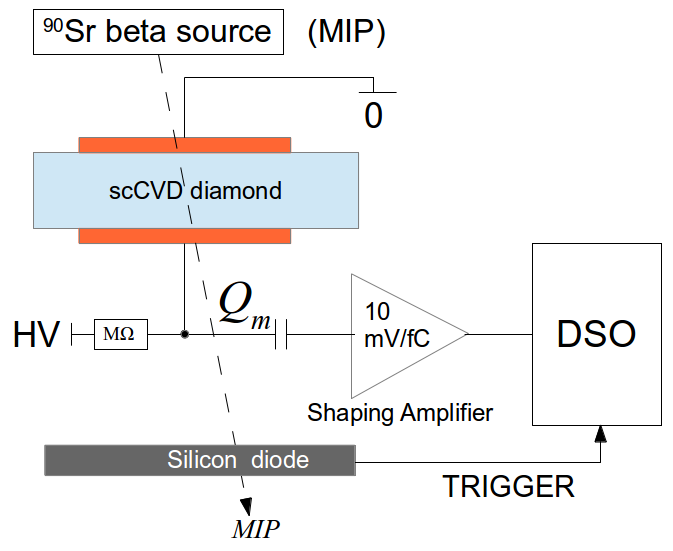}
  \caption{Setup for charge collection efficiency measurements}
  \label{fig:cce}
  \end{subfigure}
  \caption{Test setup}
\end{figure}
Figure~\ref{fig:tct} shows the setup used to perform TCT measurements. 
Bias voltage (with a current limiting M$\Omega$ resistor) is 
provided as pull-up in the signal path. The setup is arranged so that $\alpha$ 
particles enter from the opposite, grounded side of the sample. The penetration
depth of $\alpha$ in diamond is $\approx 10\upmu m$. 
The profile of current induced by primary e-h pairs as they traverse 
the bulk can be measured as a function of time.

The readout scheme for measuring the charge collection efficiency 
shown in Figure~\ref{fig:cce} is similar. A collimated 
$^{90}Sr\ \beta$ source is aligned with the diamond and a trigger 
silicon diode. A charge shaping amplifier integrates the deposited charge 
and measures $dE/dx$ in the sample.

\subsection{Results: current profile}\label{sec:tct}
\begin{figure}[!htbp]
\centering
\includegraphics[width=0.8\textwidth]
{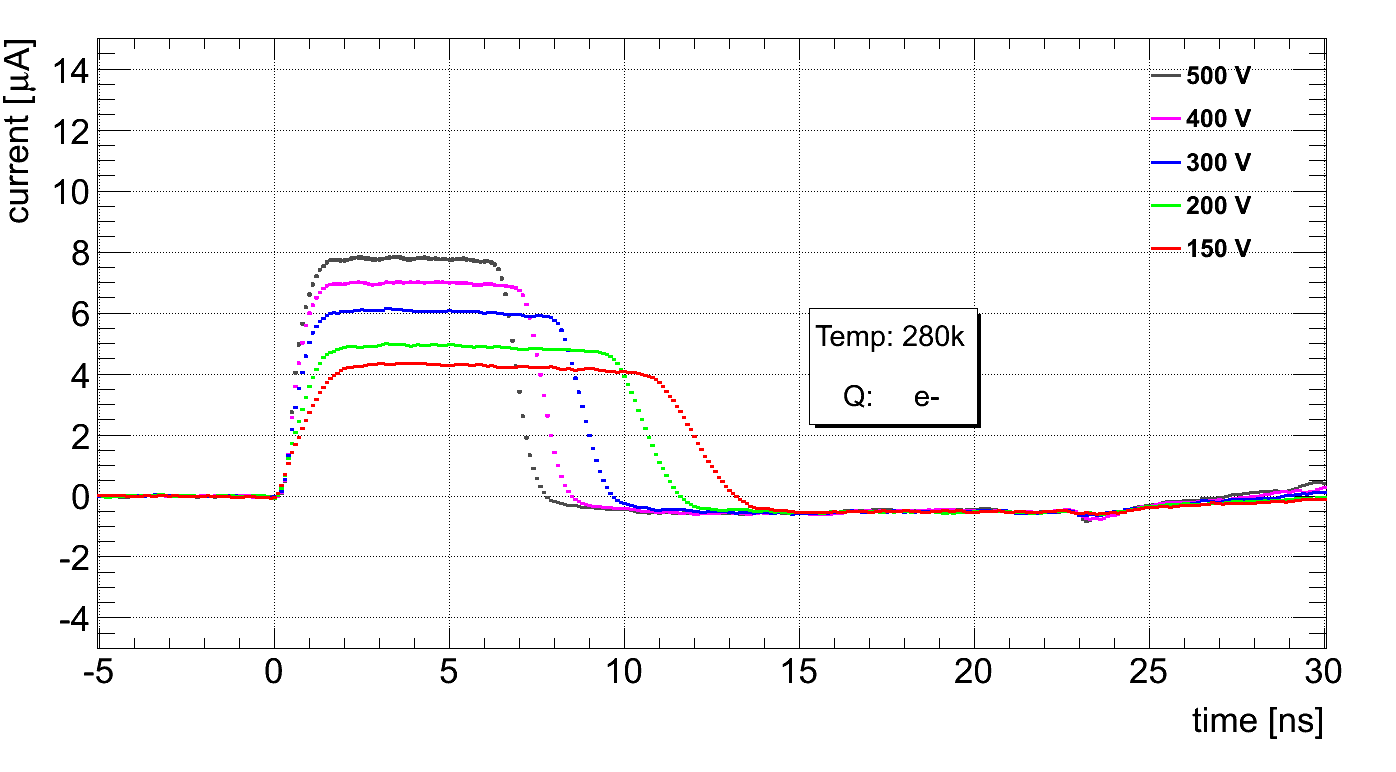}
\caption{TCT measurement of electron current at room temperature}
\label{fig:tct-rt}
\end{figure}
\begin{table}[ht]\caption{TCT results. Figures in the left column are for 
electrons, in the right column for holes}\label{tab:tct-results}
\begin{tabular}{ccc} 
280~K&
\pb{
\includegraphics[width=3.0in]{figures/TCT/ePlots/avgPulse_280k100V_e.png}
} &
\pb{
\includegraphics[width=3.0in]{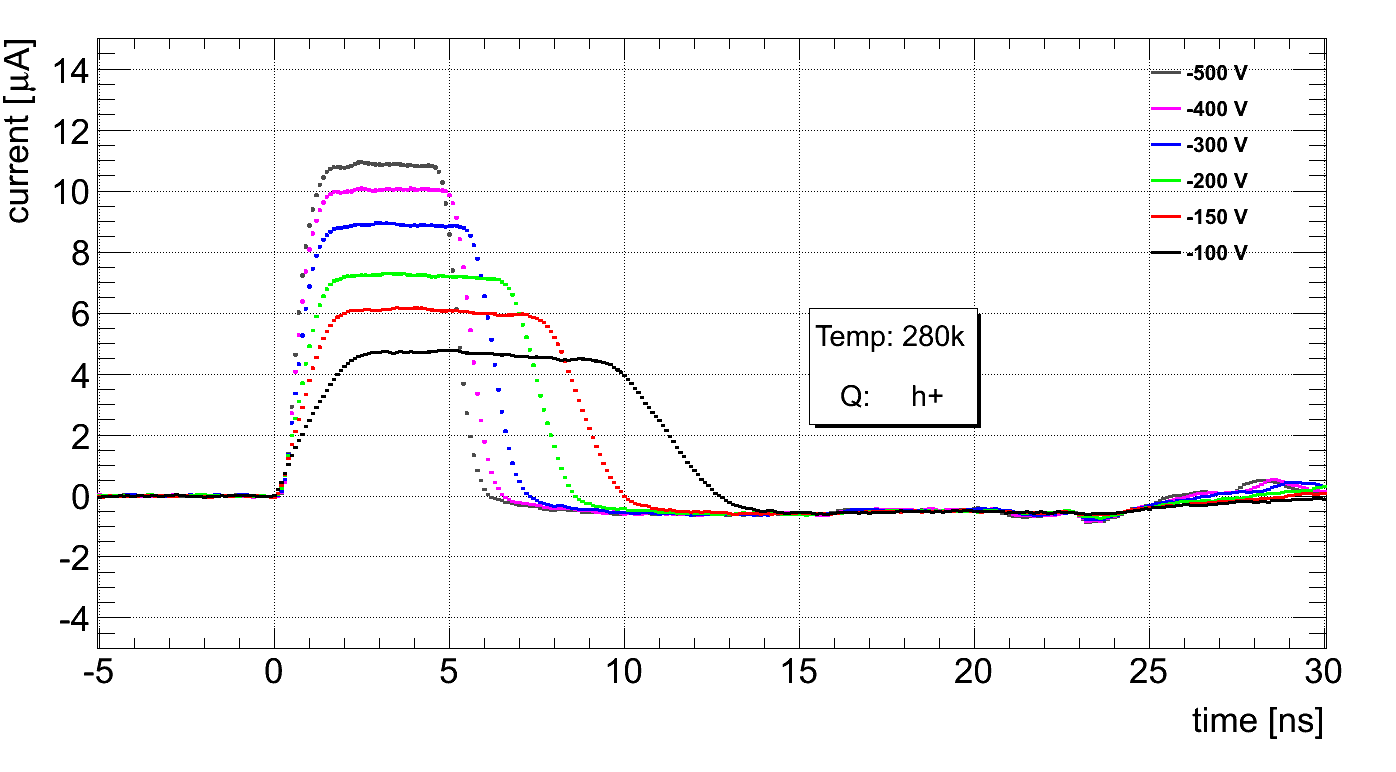}
} \\
210~K&
\pb{
 \includegraphics[width=3.0in]{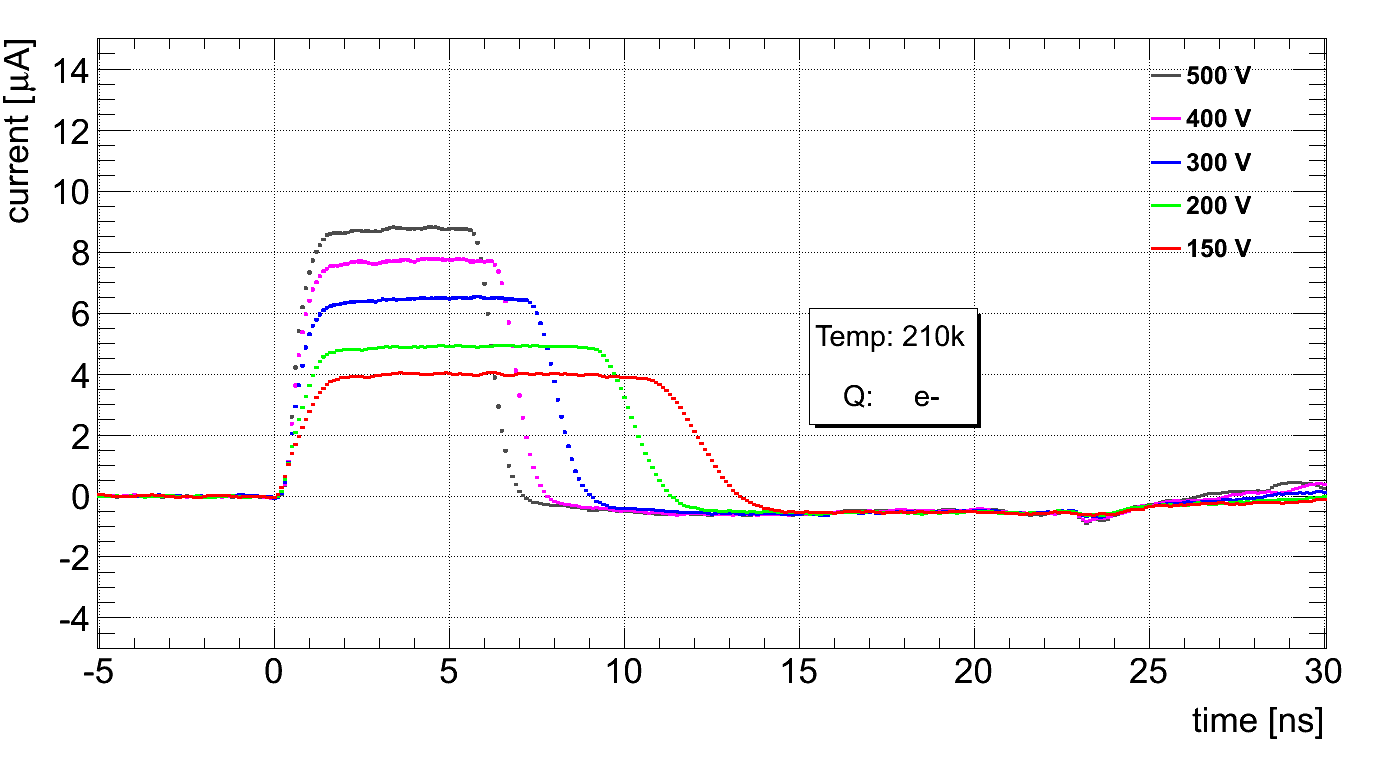}
} &
\pb{
 \includegraphics[width=3.0in]{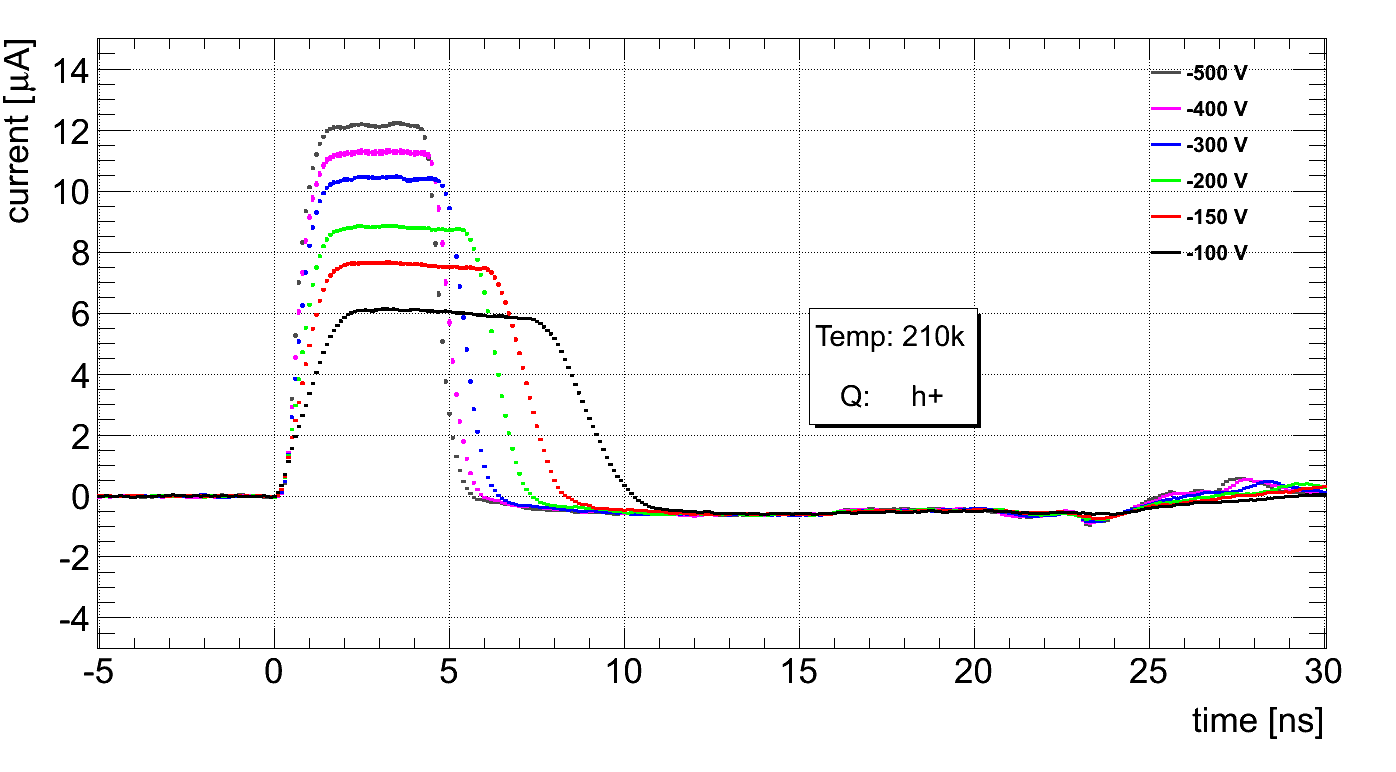}
} \\
170~K&
\pb{
 \includegraphics[width=3.0in]{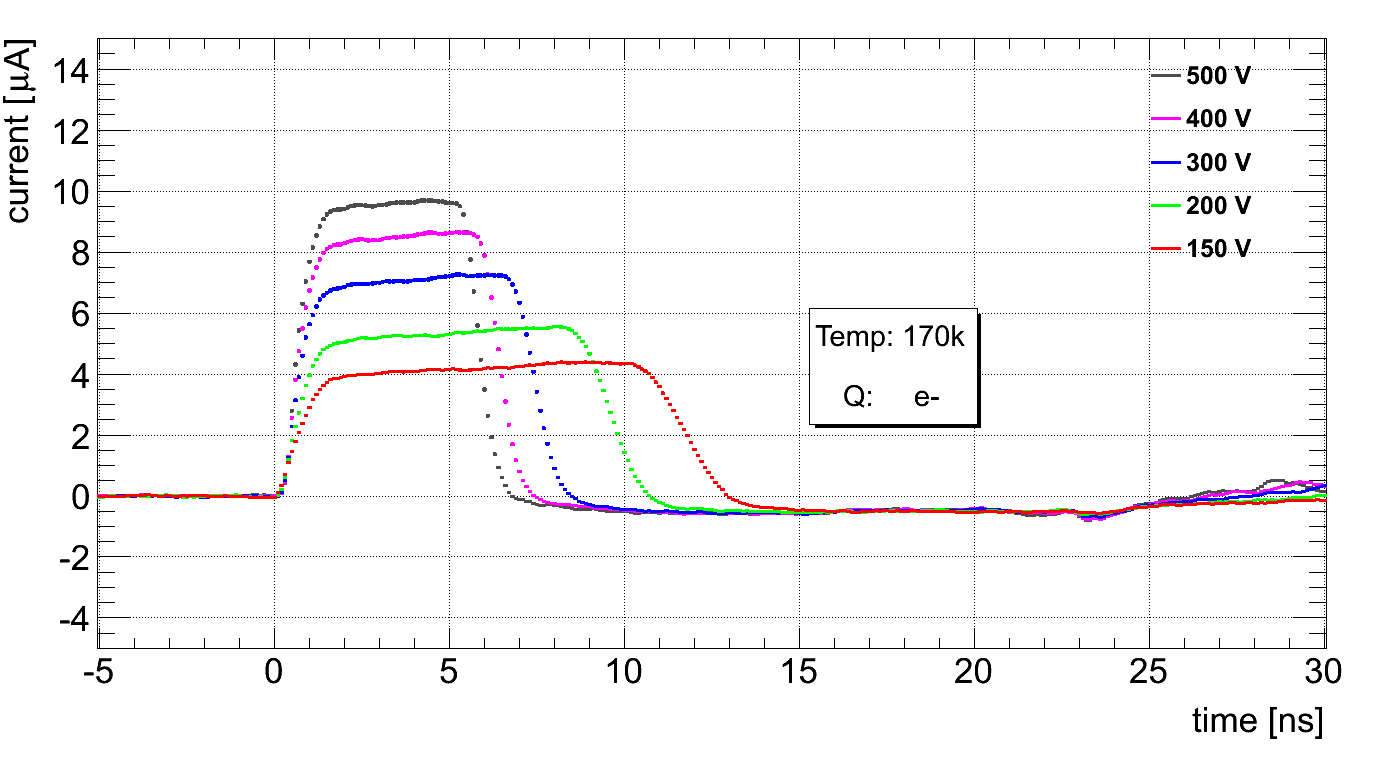}
} &
\pb{
 \includegraphics[width=3.0in]{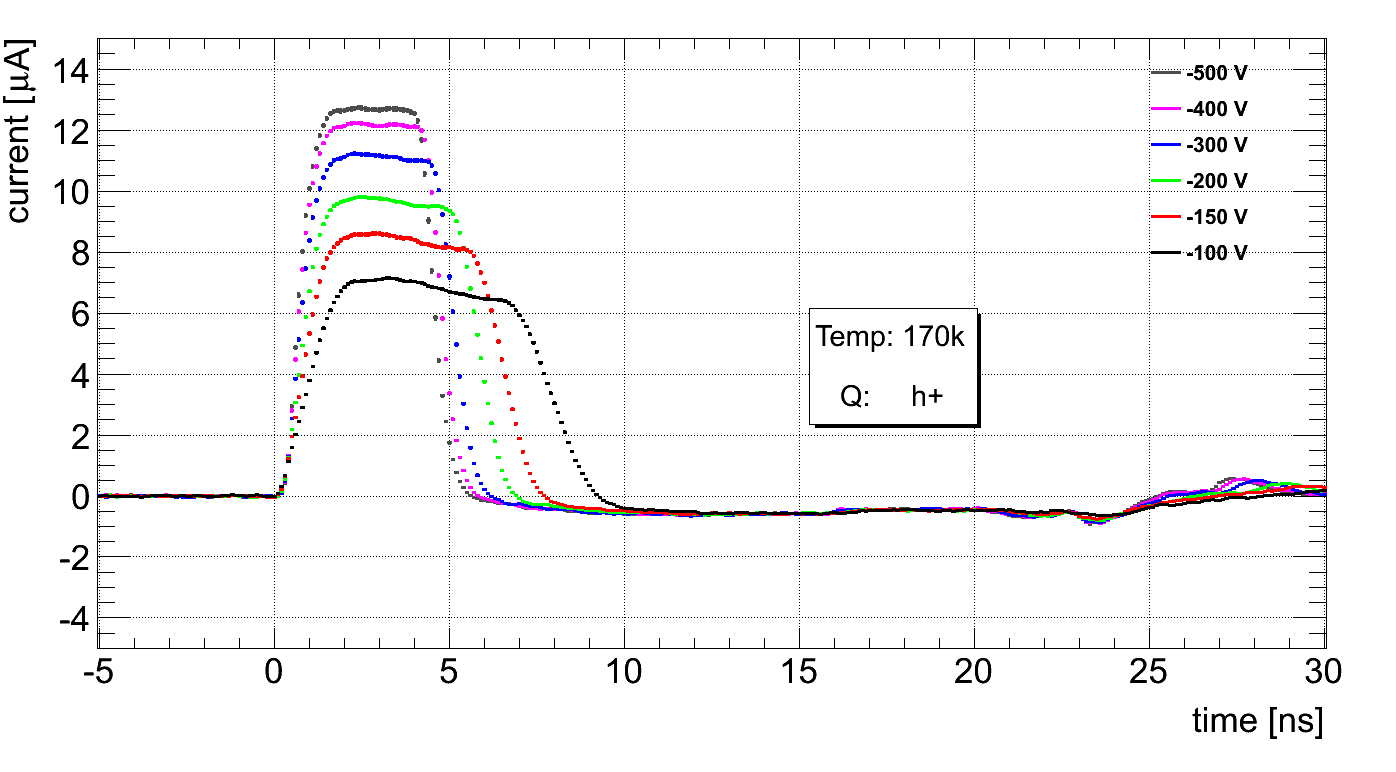}
} \\
95~K&
\pb{
 \includegraphics[width=3.0in]{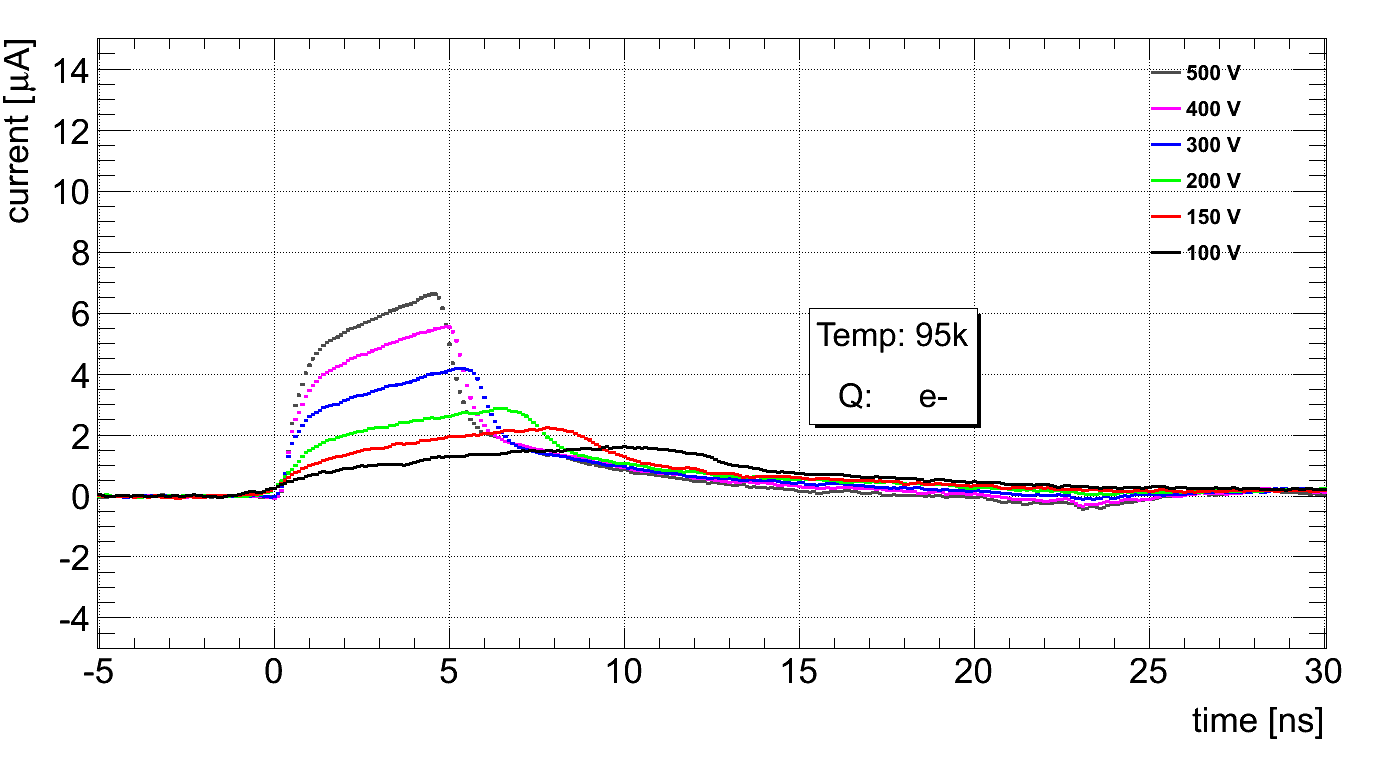}
} &
\pb{
 \includegraphics[width=3.0in]{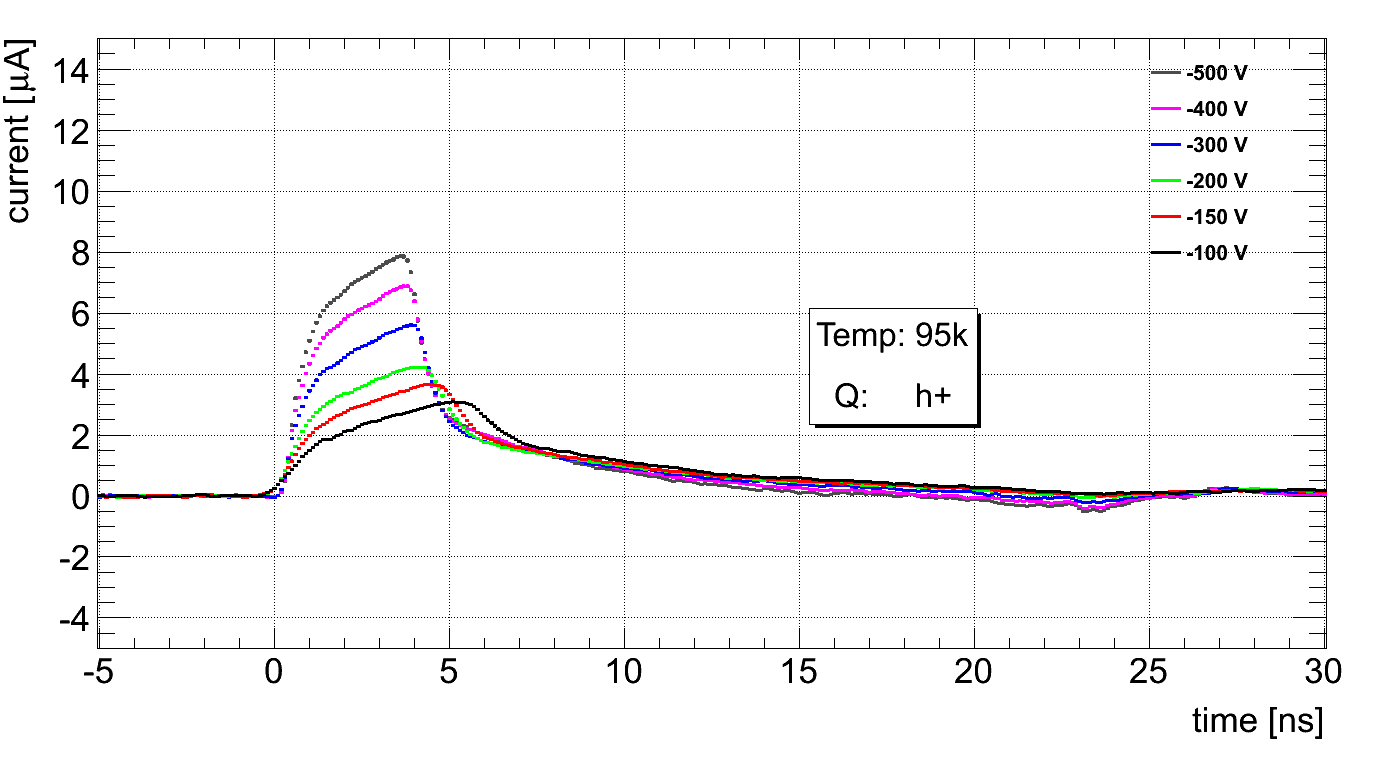}
} \\
4~K&
\pb{
 \includegraphics[width=3.0in]{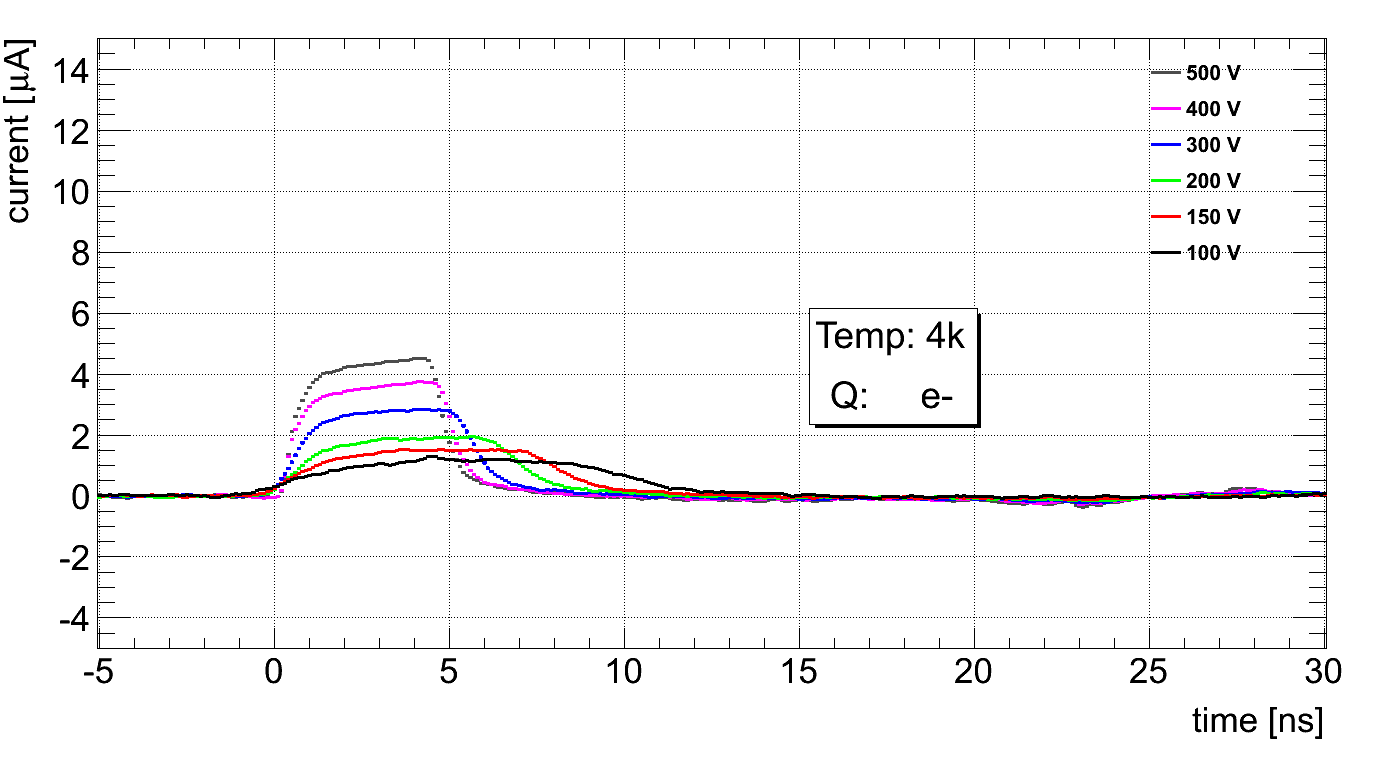}
} &
\pb{
 \includegraphics[width=3.0in]{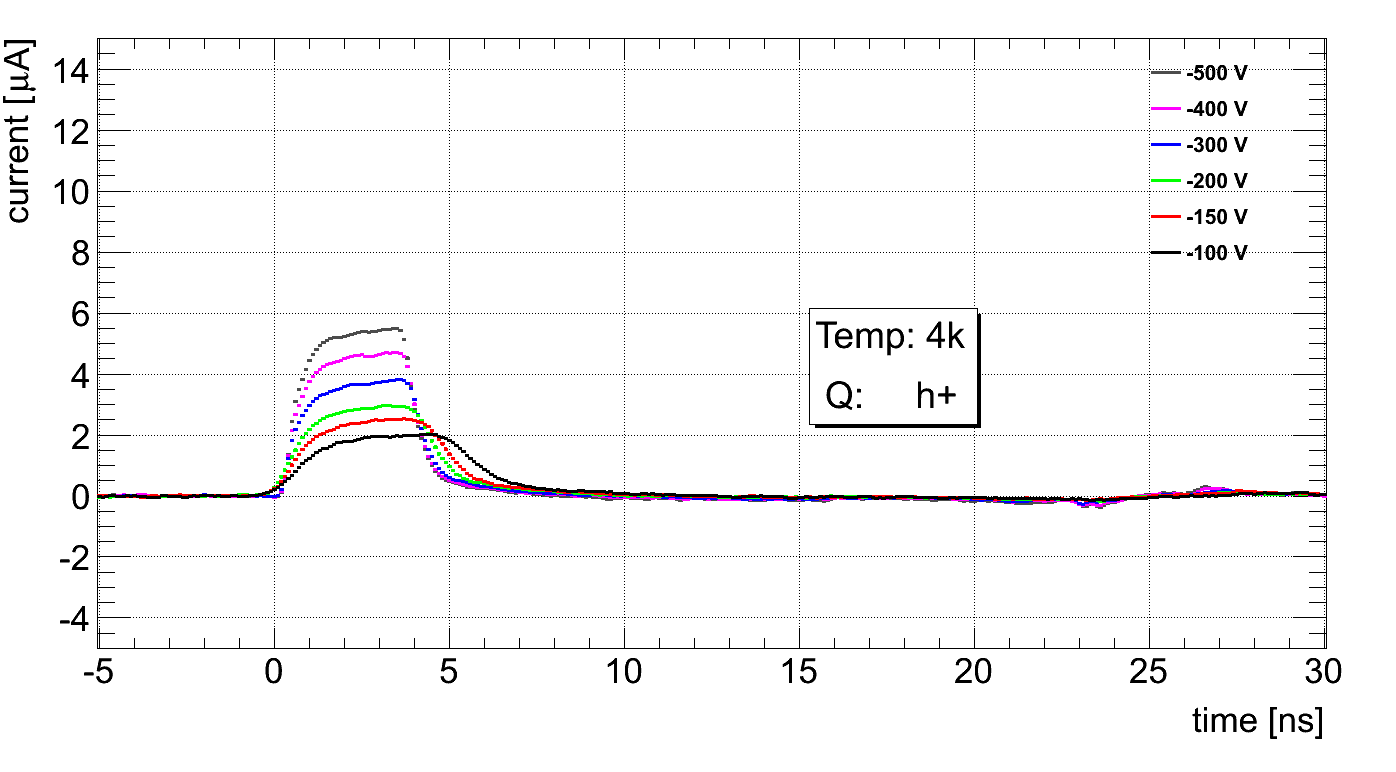}
} \\

  \end{tabular}
\end{table}
Table~\ref{tab:tct-results} shows results of the TCT current profile 
measurements as a function of bias voltage and temperature. 
Figure~\ref{fig:tct-rt} highlights the results for electrons at room 
temperature. 

The signal rises fast due to the plasma of primary e - h pairs created 
by the $\alpha$, followed by a flat phase 
during which charge carriers drift 
in the uniform electric field, inducing a constant current in the signal 
electrodes. There is an exponentially decaying tail when the charge carriers 
are collected at the electrode. At lower temperatures, there is a pronounced 
slope in the drift phase of the signal indicating the presence of
space charge. High carrier mobility and carrier recombination 
in the charge cloud at low temperature\cite{jansenref} reduces the pulse 
duration and amplitude.

\subsection{Comparison with TCAD calculations}\label{sec:tcad}
\begin{figure}[htbp]
    \centering
    \begin{subfigure}[t]{0.4\textwidth}
        \includegraphics[width=1.0\textwidth]{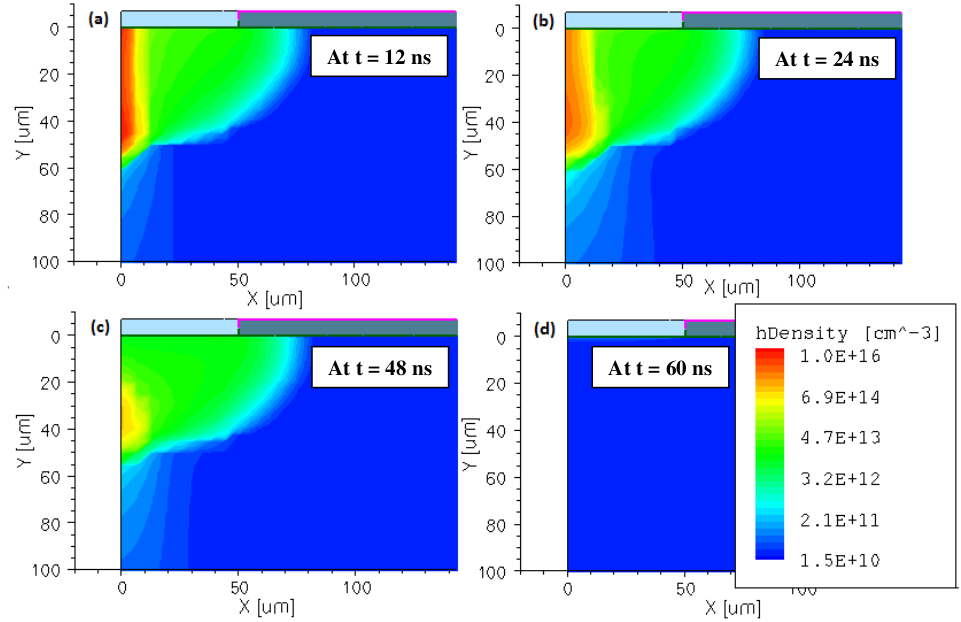}
        \caption{TCAD simulation geometry}
         \label{fig:tcad-a}
    \end{subfigure}~
    \begin{subfigure}[t]{0.5\textwidth}
        \includegraphics[width=1.0\textwidth]{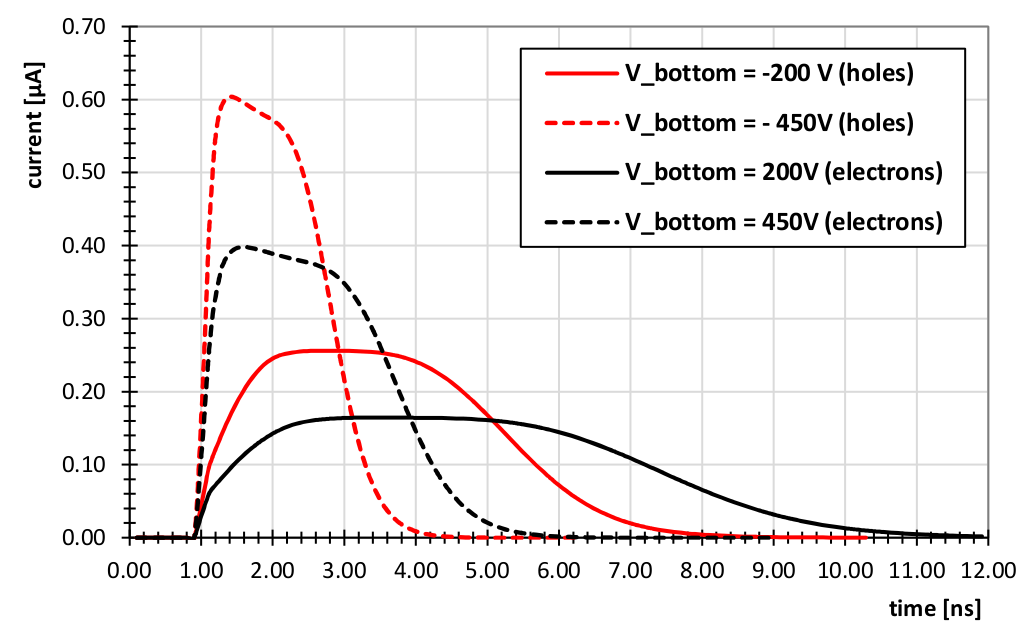}
        \caption{Results of TCAD simulation for e/h current}
        \label{fig:tcad-b}
    \end{subfigure}
   \caption{Sentaurus TCAD signal calculation}
\end{figure}

We have performed TCAD calculations in the Sentaurus tool suite to cross-check
our experimental measurements. A simplified geometry shown in 
Figure~\ref{fig:tcad-a}  with 300$\upmu m$ diamond and two dimensional 
calculation was used to improve convergence time. 
With a space charge of $3\times10^{11}cm^{-3}$ specified in
the input, the results obtained are shown in Figure~\ref{fig:tcad-b}.

The TCAD calculation results are in qualitative agreement with the trend 
of the TCT measurements in Table~\ref{tab:tct-results} but the signal magnitude 
is significantly smaller.
There are a few possible reasons for the discrepancy: a) in simulation, 
the ionizing $\alpha$ enters into the
crystal at the electrode to which high voltage is applied; b) the simplified 
two dimensional geometry and 
c) the space charge setting in the calculation. Reason (a) is
most likely since we have found that in our experimental setup 
(Figure~\ref{fig:tct}) reversing the direction of the $\alpha$ entry reduces 
the signal magnitude.

\subsection{Results: MIP signal distribution, Charge collection efficiency}\label{sec:landau}
\begin{figure}[htbp]
    \centering
    \begin{subfigure}[t!]{0.6\textwidth}
    \includegraphics[width=1.0\textwidth]{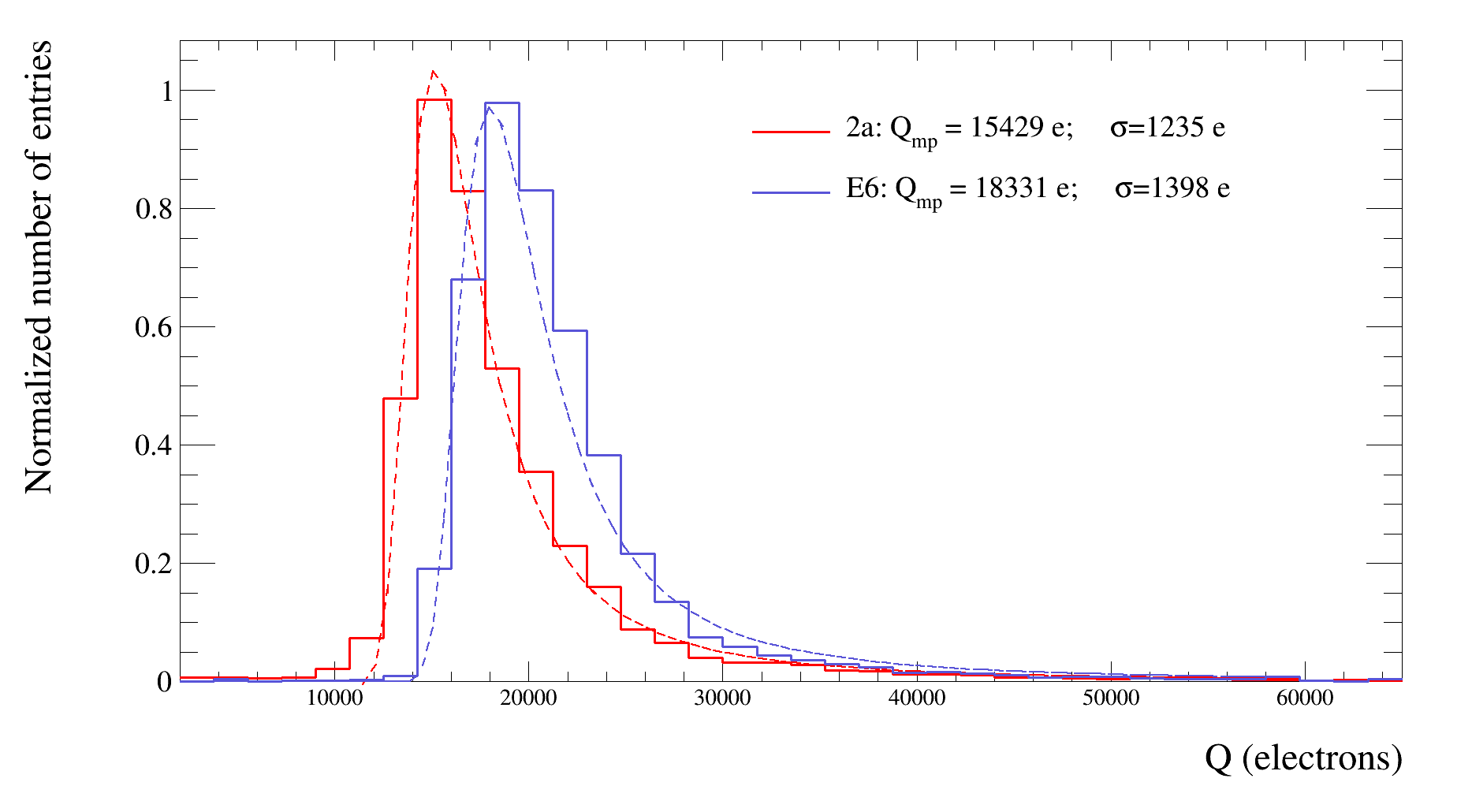}
    \caption{Signal distribution of MIP $\beta$ particles at $1V/\upmu m$ bias. $Q_{mp}$ is the
             most probable charge value obtained from the Landau fit }
    \label{fig:landau}
    \end{subfigure}~\ ~
    \begin{subfigure}[t!]{0.4\textwidth}
    \includegraphics[width=1.0\textwidth]{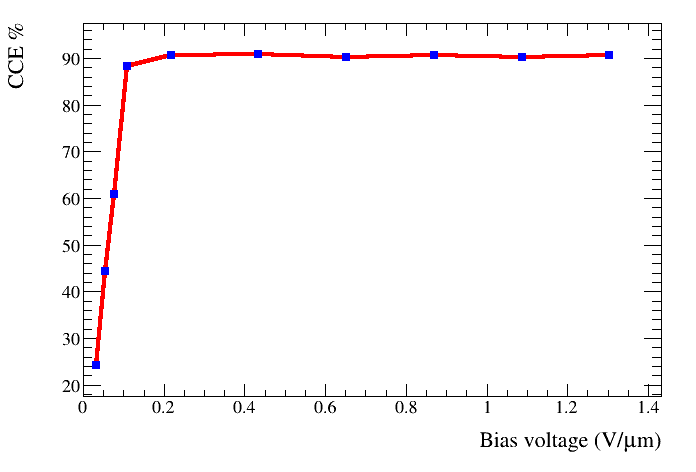}
    \caption{Charge collection efficiency in \textsc{2a} as a function of applied bias voltage}
    \label{fig:cce-bias}
    \end{subfigure}
\caption{Results of Charge collection efficiency measurements}
\vspace{0.2in}
\end{figure}

Figure~\ref{fig:landau} shows a characteristic Landau distribution of signals
recorded from passage of minimum ionizing $\beta$ particles through the diamond.
 We have shown a comparison of signals recorded in the \textsc{2a} sample and 
the \textsc{e6} sample in the same setup.

Using the benchmark figure of $\rho_{MIP}=36~eh~pairs/\upmu m$ in diamond and 
sample thickness $d=460\upmu m$ for \textsc{2a} ($d=538\upmu m$ for 
\textsc{e6}) the Charge collection efficiency (CCE) is calculated as the ratio 
of the measured charge $Q_{mp}$ to the total charge deposited in the sample:
\[ CCE\ \epsilon_Q\ =\ \frac{Q_{mp}}{\rho_{MIP}\times d}\ =\ 92.9\%\ (\textsc{2a})
;\ \ 94.5\%\ (\textsc{e6})\]
This also gives a measure of the charge collection distance 
$\delta_c=\epsilon_Q\times d$ to be  $428~\upmu m$ in \textsc{2a} and 
$509~\upmu m$ in the \textsc{e6} sample.

Our measurements of $\epsilon_Q$ as a function of the bias voltage shown in
Figure~\ref{fig:cce-bias} indicate that the sample reaches close to maximum
CCE already at 0.2V/$\upmu m$ bias.

\section{COMET: Coherent Muon to Electron detection experiment at \mbox{J-PARC}}
\begin{figure}[htbp]
    \centering
    \begin{subfigure}[t]{0.5\textwidth}
    \includegraphics[height=2in]{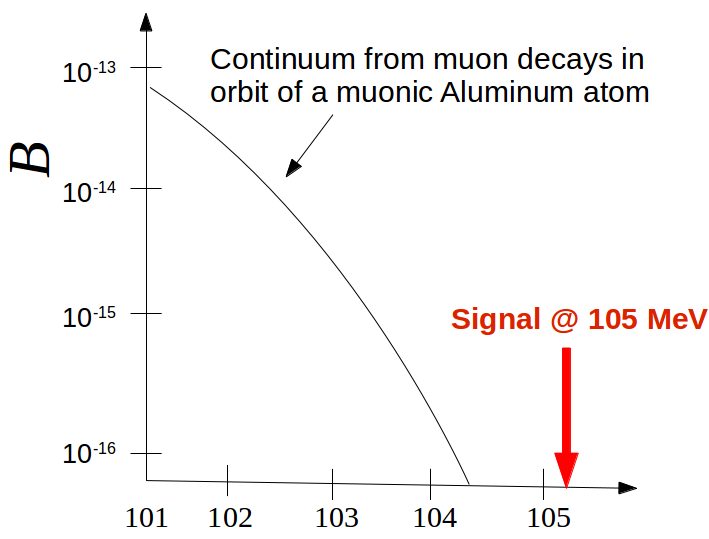}
    \caption{Sensitivity to $\upmu\rightarrow e$. $X$ axis indicates
     measured kinetic energy of the electrons in MeV}
     \label{fig:comet-signal}
     \end{subfigure}~
     \begin{subfigure}[t]{0.5\textwidth}
     \includegraphics[width=1.0\textwidth]{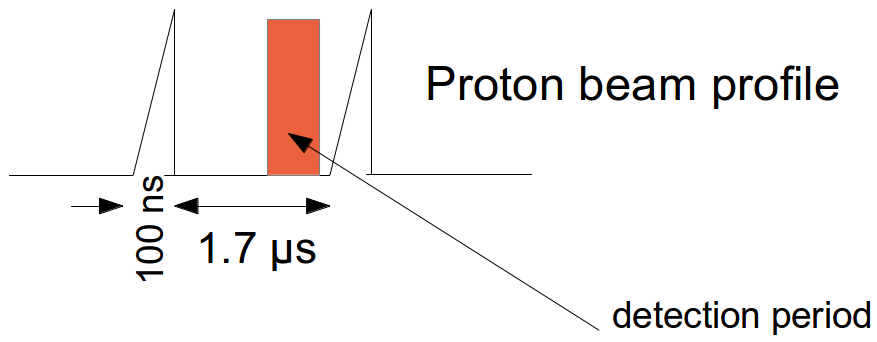}
     \caption{Time structure of the proton beam}
     \label{fig:comet-beam}
     \end{subfigure}
\caption{Signal sensitivity and beam structure in COMET}
\end{figure}

The COMET experiment at \mbox{J-PARC}\cite{cometref} will look for physics 
beyond the standard
model through charged lepton flavor violation (CLFV). 
Muons are captured in an aluminum
target to create muonic atoms. After waiting for $\approx 1.2 \upmu s$ which
allows for natural muon to electron decay, the experiment looks for the
emission of an electron with kinetic energy exactly equal to the rest mass of
the muon minus its (small) binding energy in the aluminum atom. 
This would indicate a coherent conversion of the muon to electron $-$ a CLFV 
process not allowed by the standard model. 
The expected signal stands out at the end of the continuum as shown in 
Figure~\ref{fig:comet-signal}. The sensitivity of the experiment is determined
by the muon statistics. Hence a 
beam profile monitor is necessary to ensure the purity and time structure of the beam.
\subsection{Radiation and timing requirements}
The beam structure is indicated in Figure~\ref{fig:comet-beam}. At startup, we
expect to have an 8 GeV beam of intensity 
$10^{10}$ protons/sec. 
In stable operation, the beam intensity is ramped up to $10^{14}$ protons/sec 
to increase muon flux and improve the sensitivity of the experiment.

The primary challenge in monitoring the beam profile is high proton flux 
and the necessity to ensure that the beam is completely extinguished in
100 ns. Any stray protons entering the beamline outside the 100ns bunch
 would contribute to background during the 
detection period highlighted in red in Figure~\ref{fig:comet-beam}.
The combination of high radiation dose and fast timing make
diamond an ideal choice for this application. 
\begin{figure}[!htbp]
\centering
\includegraphics[width=0.8\textwidth]{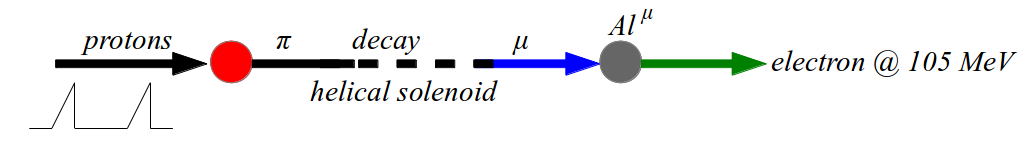}
\caption{COMET experiment layout}
\label{fig:comet}
\end{figure}

Figure~\ref{fig:comet} shows the COMET experimental setup. An 8 GeV proton beam
is slow-extracted from \mbox{J-PARC} in 100 ns bunches. The protons create pions 
in a stopping target. Pions decay in flight to muons.
 The aluminum muon capture target is placed at the end of the beamline.

\section{Conclusions}
We have measured signal properties and charge collection efficiency in scCVD
diamond from a new vendor \textsc{2a systems}. The TCT current response is excellent at
5 ns signal fwhm with no indication of space charge at room temperature, i.e.
absence of any charge trapping centers. The charge collection efficiency
in a our 460 $\upmu m$ sample at 1 $V/\upmu m$ bias is 92.9\%. Both these properties
combined with the high radiation tolerance (to be studied) make this scCVD diamond an
excellent choice to construct a beam profile monitor for the COMET experiment 
at \mbox{J-PARC}.

\acknowledgments
We are grateful to the following for help and support:
\textsc{2a Systems}, Singapore for providing the scCVD diamond;
Cividec GmbH for the readout amplifiers,
CERN PH-ADE-ID and the CERN CRG group for access to cryogenic test
facilities, IIT Bombay Industrial Research and Consultancy Center (MHRD, Government of India) for travel support and
IIT Bombay Department of Electrical Engineering, for access to Sentaurus 
TCAD.

\end{document}